\documentstyle[twocolumn,aps,epsfig,graphics]{revtex}
\begin{document}

\title{Multi-strange baryon production in Au+Au collisions near threshold}

\author{Gebhard Zeeb, Manuel Reiter, and  Marcus Bleicher \vspace*{.5cm}}

\address{ Institut f\"ur Theoretische Physik, \\
J. W. Goethe-Universit\"at, \\
Robert-Mayer-Str.\ 8-10, 60054 Frankfurt am Main, Germany}

\maketitle
{\sf The centrality dependence of $\Xi^-$ and $\Lambda$ production
in Au+Au interactions at $E_{\rm lab}=6$~$A$GeV is studied 
within a microscopic transport approach.
In line with recent data, a slight enhancement of the $\Xi^-/(\Lambda+\Sigma^0)$ 
ratio toward central collisions is found. It is demonstrated that the observed
production of multiple strange baryons can be traced back to multi-step meson-baryon 
interactions in the late stages of the collisions. Therefore, the present
analysis supports an interpretation of the observed  $\Xi$ abundance in terms of
hadronic re-scattering.}
\vspace{.6cm}

The major goal of the various heavy ion programs is the search for
a transient state of deconfined matter, dubbed the quark-gluon-plasma (QGP):
A phase transition to this new state of matter is predicted by lattice
QCD when a sufficiently high energy density
($\epsilon \approx 1$~GeV/fm$^3$) is reached \cite{qgpreviews}.

Strange particle yields and spectra are key probes
to study excited nuclear matter and to detect
the transition of (confined) hadronic matter to
quark-gluon-matter, i.e.~QGP\cite{qgpreviews,rafelski}.
The relative enhancement of strange and multi-strange
hadrons, as well as hadron ratios in central
heavy ion collisions with respect to peripheral or proton
induced interactions have been suggested as a signature
for the transient existence of a QGP phase \cite{rafelski}.

A wealth of systematic information has been gathered to study the
energy dependence of observables from
$\sqrt s \approx 2$~$A$GeV to $\sqrt s=200$~$A$GeV.
For the first time also information on the $\Xi$ near threshold 
is available \cite{Chung:2003zr}.

For our investigation, the Ultra-relativistic Quantum Molecular
Dynamics model (UrQMD 1.2)~\cite{urqmd} is applied
to heavy ion reactions from $E_{\rm lab}=2$~$A$GeV to 10~$A$GeV.
This microscopic transport approach is based on the covariant propagation of
constituent quarks and diquarks accompanied by mesonic and baryonic
degrees of freedom. It describes multiple interactions of
ingoing and newly produced particles, the excitation
and fragmentation of color strings and the formation and decay of
hadronic resonances.
Toward higher energies, the treatment of sub-hadronic degrees of freedom is
of major importance.
In the present model, these degrees of freedom enter via
the introduction of a formation time for hadrons produced in the
fragmentation of strings \cite{andersson87a,andersson87b,sjoestrand94a}.
The leading hadrons of the fragmenting strings contain the valence-quarks
of the original excited hadron. In UrQMD they are allowed to
interact even during their formation time, with a reduced cross section
defined by the additive quark model,
thus accounting for the original valence quarks contained in that
hadron \cite{urqmd}. A phase transition to a quark-gluon state is
not incorporated explicitly into the model dynamics.
However, a detailed analysis of the model in thermal equilibrium
yields an effective equation of state of
Hagedorn type \cite{Belkacem:1998gy,Bravina:1999dh}.

Note that the present calculations have been performed without
potential interactions. Thus, no in-medium modifications of hadron 
masses and widths are taken into account.

\begin{figure}
\resizebox*{!}{0.5\textheight}{\includegraphics{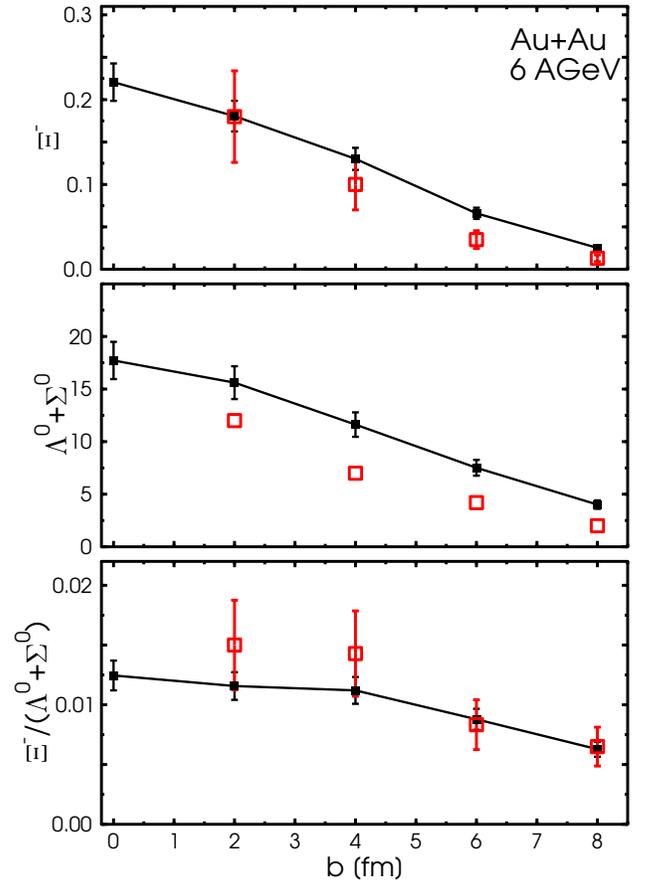}} 
\vspace*{.3cm}
\caption{Centrality dependence of the ratio $\Xi^-/(\Lambda+\Sigma^0)$ 
(bottom) and the 
$\Lambda$ (middle) and $\Xi^-$ (top) yields 
in  Au+Au interactions at $E_{\rm lab}=6$~$A$GeV.
Small symbols denote the calculations, while large symbols
show the data from \protect\cite{Chung:2003zr}.
\label{bdepall}}
\end{figure}

Figure~\ref{bdepall} (top and middle panel) shows the
centrality dependence of the $\Xi^-$ and $\Lambda$ yields 
in Au+Au interactions at $E_{\rm lab}=6$~$A$GeV.
%%%%%%%%%%%%%%%%%%%%%%%%%%%%%%%%%%%%%%%%%%%%%%%%%%%%%%%%%%%%%%%%%%%%%
%%%%%%%%%%%%%%%%%%%%%%%%%%%%%%%%%%%%%%%%%%%%%%%%%%%%%%%%%%%%%%%%%%%%

One clearly observes a strong increase of the \mbox{(multi-)}strange baryon
yields toward head-on collisions.
However, at all energies from 2-10~$A$GeV, the
Cascade yield increases stronger than the single strange 
hyperon abundance toward central collisions. 
This leads to a moderate centrality dependence of the $\Xi^-/\Lambda$ ratio
as shown in Fig.~\ref{bdepall} (bottom) explicitly for a beam
energy of 6~$A$GeV. 
The centrality dependent yields (small symbols show the calculations, 
while large symbols denote the experimental data 
taken from \cite{Chung:2003zr}) as well as the ratios show good
agreement with the recently measured data of the E895 collaboration.

Let us now focus on the dynamics of the Cascade production. 
In the present approach the increase in strangeness 
production towards central interactions is due to multi-step processes. 
I.e., Primary interactions lead to the production of
\begin{enumerate}
\item
heavy mesons and baryons 
\item
as well as $\Lambda$ and $\Sigma$ hyperons and anti-Kaons.
\end{enumerate} 
In subsequent interactions, the collisions of those excited meson/baryon 
resonances can more easily result in the production of a Cascade,
due to their increased center of mass energy.
In Ref.\ \cite{Spieles:1993jx}, a similar mechanism has been 
advocated to understand the production of anti-baryons around 
threshold energies.
In addition, meson interactions with previously formed hyperons
have also an increased chance for $\Xi$ production, because only
one additional unit of strangeness has to be produced.
Finally strangeness exchange reactions, 
e.g.\ $\overline K + \Lambda \rightarrow \Xi+\pi$, can have a major
contribution to the finally observed Cascade abundance.

Three processes and classes of channels for $\Xi$ production can be
identified:
\begin{itemize}
\item
Reactions between baryons (BB), i.e. BB $\rightarrow \Xi +x$.
\item
Meson-baryon reactions (MB), i.e. MB $\to \Xi +x$.
\item 
Decays of baryon resonances, i.e. $\Xi^*\rightarrow \Xi +x$.
\end{itemize}
Note that $\Xi$ includes also $\Xi^*(1530)$, since the 
$\Xi^*(1530)$ decays to 100\% into $\Xi$. Throughout this paper, 
$\Xi^*$ denotes all Cascade resonances, except the $\Xi^*(1530)$.

\begin{figure}
\resizebox*{!}{0.39\textheight}{\includegraphics{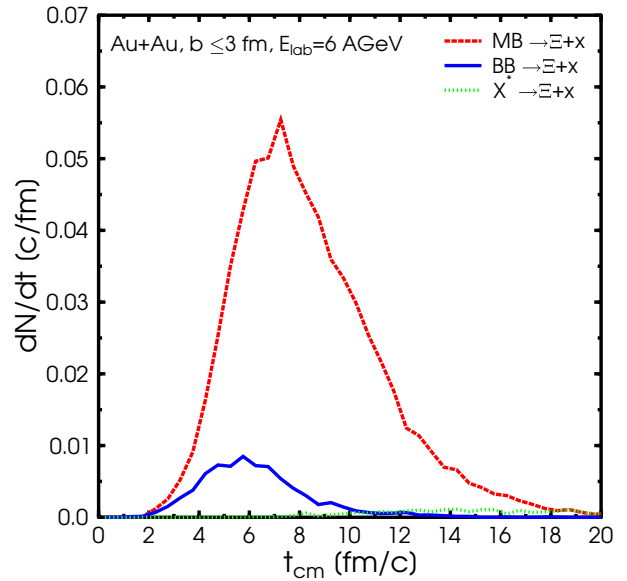}} 
%\vspace*{.3cm}
\caption{Differential production times of $\Xi$ particles in central 
Au+Au reactions at $E_{\rm lab}=6$~$A$GeV broken down to the different 
channels.
\label{xir0}}
\end{figure}

In the following, we will investigate these production channels
in more detail. 
First, it is demonstrated that  Cascades are formed
in the late stages of central heavy ion interactions at AGS energies.
Figure~\ref{xir0} shows the $\Xi$ production rate as a function of time
in central Au+Au reactions at $6~A$ GeV divided into the contribution
from baryon-baryon (BB), meson-baryon (MB) interactions and decay channels.
A clear separation in time shows up when comparing the 
different channels:
The baryon-baryon channel (dashed line) has its production 
peak at maximum overlap time of the penetrating nuclei 
at $t_{\rm max, BB} \approx$ 6 fm/c.
Then, $\Xi$ production from the cooking meson-baryon ''soup'' sets 
in (full line).
Here the maximum is at times around $t_{\rm max, MB} \approx$ 8 fm/c.
Up to 13 fm/c the production rate in the meson-baryon channel
is higher than the maximum production rate in baryon-baryon 
collisions. This clearly indicates the importance of secondary 
interaction processes for multi-strange baryon production in 
this energy range. The late contribution - from the decay of high mass  
$\Xi$ resonances (dotted line) - is tiny. 
The production of Cascades % $\Xi$ particles 
stretches over nearly 10 fm/c.
Thus, multi-strange baryon production relies on long living hadronic
stages, which might allow for  chemical equilibration even in this exotic
channel.
%%%%%%%%%%%%%%%%%%%%%%%%%%%%%%%%%%%%%%%%%%%%%%%%%%%%%%%%%%%%%%%%%%%%%%%%5
%%%%%%%%%%%%%%%%%%%%%%%%%%%%%%%%%%%%%%%%%%%%%%%%%%%%%%%%%%%%%%%%%%%%%

It is noteworthy that also other scenarios for an enhanced
production of multi-strange baryons are possible.
E.g.\ a mass reduction of the $\Xi$ might yield considerably
larger contributions from baryon-baryon scatterings, since
at the time of complete overlap baryon densities are highest.
It is hard to distinguish between our scenario (without in-medium 
modifications) and a calculation with in-medium effects included
solely on the basis of total multiplicities.
A detailed study of the in-plane and out-of-plane flow
might allow deeper insight into the underlying 
production mechanisms.

\begin{figure}
%\vspace*{-.9cm}
\resizebox*{!}{0.39\textheight}{\includegraphics{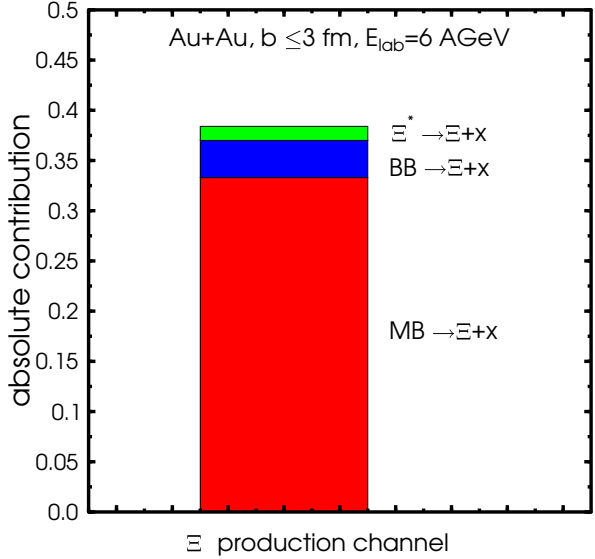}} 
%\vspace*{.3cm}
\caption{Channel decomposition of reactions leading to the production of 
$\Xi$ particles in central Au+Au reactions at $E_{\rm lab}=6$~$A$GeV.
\label{xicomp}}
\end{figure}

Figure \ref{xicomp} gives the production channel decomposition 
of Cascades in central Au+Au reactions at $E_{\rm lab}=6$ $A$ GeV. 
The leading production channel is from 
meson-baryon reactions, followed by the baryon-baryon channel. The
decay of higher mass resonances into a $\Xi$ is only of minor importance.
Meson-meson interactions do not contribute to the $\Xi$ production
due to their small number and low center of mass energies.
%%%%%%%%%%%%%%%%%%%%%%%%%%%%%%%%%%%%%%%%%%%%%%%%%%%%%%%%%%%%%%%%%%%%%%
%%%%%%%%%%%%%%%%%%%%%%%%%%%%%%%%%%%%%%%%%%%%&&&&&&&&&&&&&&&&&&&&&&&&&&
%%%%%%%%%%%%%%%%%%%%%%%%%%%%%%%%%%%%%%%%%%%%%%%%%%%%%%%%%%%%%%%%%%%%%%
\begin{figure}
%\vspace*{-.8cm}
\resizebox*{!}{0.39\textheight}{\includegraphics{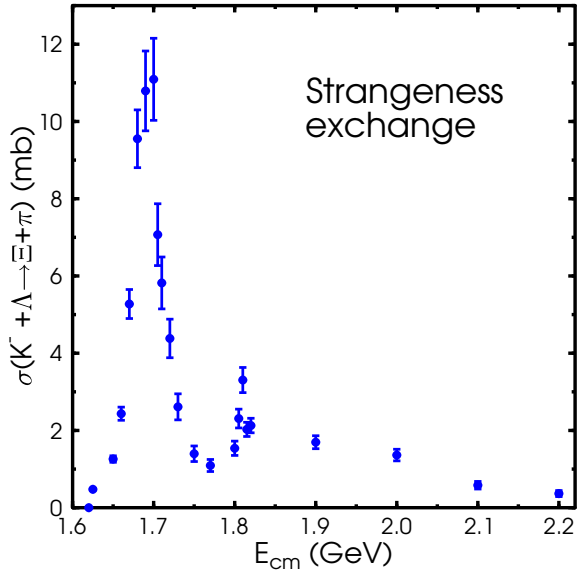}} 
%\vspace*{.3cm}
\caption{Exclusive cross section for the strangeness exchange
in ${\rm K}^-+\Lambda \rightarrow \Xi+\pi$ as a function of
the center of mass energy.
\label{kl}}
\end{figure}
%%%%%%%%%%%%%%%%%%%%%%%%%%%%%%%%%%%%%%%%%%%%&&&&&&&&&&&&&&&&&&&&&&&&&&

Let us have a deeper look into the leading production channels.
It has been suggested in \cite{Pal:2002hg} (within the ART transport 
model) that strangeness exchange reactions, 
e.g. $\overline{\rm K} (\Lambda,\Sigma) \leftrightarrow \Xi \pi$, 
are the most prominent $\Xi$ creation channels. However, the present 
study indicates that also OZI forbidden channels yield
a huge contribution to the Cascade production cross section.
To make the present work comparable to the analysis by \cite{Pal:2002hg}, 
Fig.~\ref{kl} gives the cross section for the 
reaction ${\rm K}^-+\Lambda \rightarrow \Xi+\pi$ as employed
in the present calculation. 
However, note that the presented strangeness exchange channel is
implicitely given by the baryon resonances and string break-up mechanism  
included in the model.
Possible reasons for the differences between the models are
({\bf I}) the omission of high mass meson and baryon resonances in the ART model
and ({\bf II}) a more than twice as large $\overline{\rm K}+\Lambda \rightarrow \Xi+\pi$ 
cross section in the calculation by \cite{Pal:2002hg} compared to our cross section.

The contribution of the baryon-baryon collisions to the $\Xi$ production 
is dominated by $N +\Lambda$ (42\%) and $(\Delta, N^*) + \Lambda$ (48\%),
here $\Delta, N^*$ denotes all Delta and $N^*$ resonances. 
The missing 10\% of BB interactions are the contributions of 
all remaining types of BB interactions.
In the meson-baryon sector, three distinct groups of reactions
dominate the $\Xi$ production at the investigated energy as shown
in table \ref{tablemm}. The channels $\eta$, $\eta'$, $a_0(980)$,
$a_1(1260)$, $a_2(1320)$ and $b_1(1235)$ interacting with a baryon, 
do contribute with 16\%.
\begin{table}
\begin{tabular}{l|r}
Channel  & Percentage \\\hline\hline
($\overline K, \overline K^*(892)$) + B & 33 \% \\\hline
$\pi$ + B  & 24 \% \\\hline
($\rho, \omega$) + B & 22 \% \\\hline
($\eta$, $\dots$, $b_1(1235)$) + B & 16 \%\\ \hline
other MB reactions   & 5 \% \\\hline\hline
Total                & 100 \%
\end{tabular}
\caption{\label{tablemm} Decomposition of the dominant meson induced
channels to the $\Xi$ production.}
\end{table}
\vspace*{-.3cm}
\begin{table}
\begin{tabular}{l|r|r}
Reaction & rel. & abs. \\ \hline\hline
($\overline K, \overline K^*(892)$)  + N       & 11 \% & 3.7 \%  \\\hline
($\overline K, \overline K^*(892)$)  + ${\rm N}^*$    & 21 \% & 6.9 \%  \\\hline
($\overline K, \overline K^*(892)$)  + $\Delta$ & 34 \% & 11.2 \% \\\hline
($\overline K, \overline K^*(892)$)  + $\Lambda,\Sigma$& 34 \% & 11.2 \% \\\hline
Total                                                  & 100 \% & 33.0 \%\\\hline\hline
$\pi$ + ($\Lambda,\Sigma$) & 91 \% & 21.8 \% \\\hline
$\pi$ + other baryon       & 9 \%  & 2.2 \% \\ \hline           
Total                      & 100 \% & 24.0 \% \\\hline\hline
($\rho$, $\omega$) + ($\Lambda,\Sigma$) & 98 \% & 21.6 \% \\\hline
($\rho$, $\omega$) + other baryon       & 2 \%   & 0.4 \% \\ \hline
Total                                   & 100 \% & 22.0 \%\\\hline\hline
($\eta$, $\dots$, $b_1(1235)$) + N       & 14 \% & 2.2 \% \\\hline
($\eta$, $\dots$, $b_1(1235)$) + $\Delta$& 7 \%  & 1.1 \% \\\hline
($\eta$, $\dots$, $b_1(1235)$) + $\Lambda,\Sigma$& 76 \% & 12.2 \% \\\hline
($\eta$, $\dots$, $b_1(1235)$) + other baryon & 3 \% & 0.5 \% \\ \hline
Total                                   & 100 \% & 16 \%
\end{tabular}
\caption{\label{tableb} Classification of the baryon types 
in meson induced reactions leading to the production of a Cascade.}
\end{table}
%%%%%%%%%%%%%%%%%%%%%%%%%%%%%%%%%%%%%%%%%%%%%%%%%%%%%%%%%%%%%%%%%%%5
\begin{figure}
\resizebox*{!}{0.39\textheight}{\includegraphics{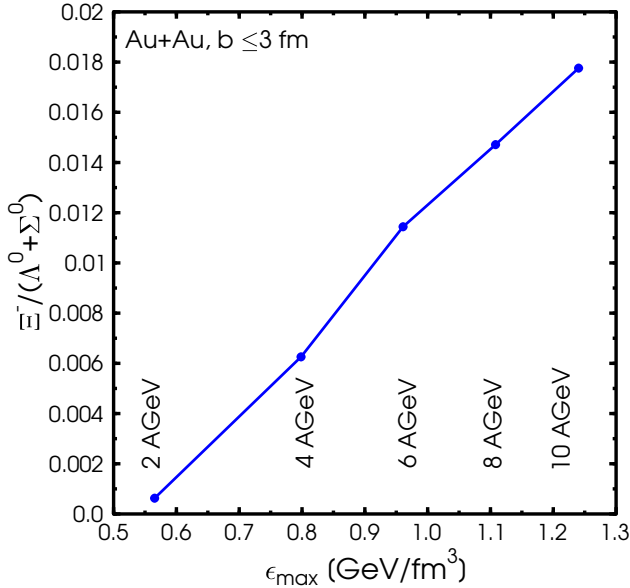}}
%\vspace*{.3cm}
\caption{The ratio $\Xi^-/(\Lambda+\Sigma^0)$ in central
Au+Au reactions as a function of the maximal
energy density achieved at incident energies from 2~$A$GeV to 10~$A$GeV.
\label{erhodep}}
\end{figure}
%%%%%%%%%%%%%%%%%%%%%%%%%%%%%%%%%%%%%%%%%%%%%%%%%%%%%%%%%%%%%%%%%%%%

The Kaon channel is dominated by
the $\overline K$ with 59\% of all $\overline K$ induced MB reactions 
and $\overline K^*(892)$ (27\%). The six high mass anti-Kaon resonances
do each contribute 0.6\% - 4\%.
The different meson induced channels are given explicitely
in table \ref{tableb}.

From the detailed analysis presented, it is evident that $\Xi$ production
is mainly driven by meson-baryon interactions including a $\Lambda$ or $\Sigma$.
In fact, 70\% of all meson-baryon reactions resulting in the
production of a  Cascade include a hyperon.
However, the present analysis indicates that only 11\% of the MB interactions
leading to $\Xi$ production proceed via the strangeness exchange reaction 
anti-Kaon + hyperon, in strong contrast to Ref.\ \cite{Pal:2002hg}. 

Why is the energy range of $E_{\rm lab}=2$ to 10~$A$GeV so 
interesting and important? This becomes obvious by studying the maximal
achieved energy density in this energy region.
Figure~\ref{erhodep} shows the ratio $\Xi^-/\Lambda$ in central
Au+Au reactions as a function of maximal energy density reached
in collision at different incident energies. At a beam energy of 6~$A$GeV,
the model predicts to cross the critical energy density of 1~GeV/fm$^3$.
The present calculation predicts a linear dependence of $\Xi^-/(\Lambda+\Sigma^0)$
on the energy density. Therefore, it is important to study deviations from
this scaling to identify a possible onset of additional
strangeness enhancement from non-hadronic sources.

\begin{figure}
%\vspace*{-.4cm}
\resizebox*{!}{0.39\textheight}{\includegraphics{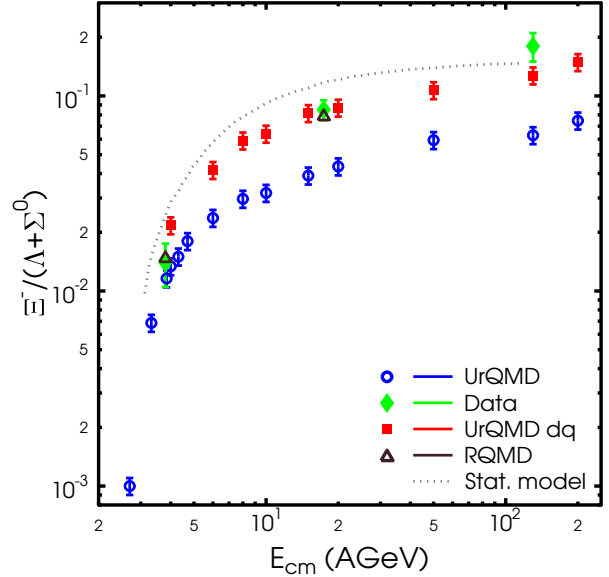}} 
%\vspace*{.3cm}
\caption{Energy dependence of the $\Xi^-/(\Lambda+\Sigma^0)$ ratio
in Au+Au/Pb+Pb collisions.
The UrQMD calculations ($b=2$~fm) are shown by circles. UrQMD with the inclusion
of di-quark clusters is depicted by the squares. The dotted line indicates
the results of a statistical model\protect\cite{Braun-Munzinger:2001as},
while the triangles show RQMD results.
The data (diamonds) are taken from \protect\cite{data}.
\label{exc}}
\end{figure}

Finally, we compare the model calculations to $\Xi^-/(\Lambda+\Sigma^0)$ ratios
obtained at higher energies. Figure \ref{exc} shows the default UrQMD
simulation (open circles), 
%%%%%%%%%%%%%%%%%%%%%%%%%%%%%%%%%%%%%%%%%%%%%%%%%%%%%%%%%%%%%%%%%%%%%%
%%%%%%%%%%%%%%%%%%%%%%%%%%%%%%%%%%%%%%%%%%%%&&&&&&&&&&&&&&&&&&&&&&&&&&
\noindent
a calculation with strange di-quark 
clustering\footnote{This option is available with the input 
parameter cto 37 set to 1.} enabled (open squares) in comparison to 
a statistical model prediction \cite{Braun-Munzinger:2001as} (line)
and the available data \cite{data} (full diamonds).
In the AGS energy range, the hadronic dynamics are clearly able
to describe the data. Also, the  microscopic simulation and 
the statistical model agree well at low energies.
With increasing center of mass energy, however, the default simulation
clearly underpredicts the yield of double strange baryons,
resulting in an underprediction of the $\Xi^-/(\Lambda+\Sigma^0)$
ratio at SPS and RHIC. It is possible to obtain a
reasonable description of the $\Xi^-$ abundances, if 
additional mechanisms like di-quark clustering (employed in this 
investigation) or string fusion
(see Ref.\ \cite{Sorge:1995dp,Sorge:1997qw,Soff:1999et}, 
RQMD results are depicted by triangles). 

In summary, we have used a relativistic hadronic transport model to
investigate the production of multi-strange baryons 
in high density nuclear matter.
Comparing the calculation with the recent data
by the E895 collaboration indicates a good description of the data
within the present hadronic model.
In fact the nice agreement between experimentally observed yield 
and the present calculation in the AGS 
energy range is in  stark contrast to the results 
obtained at SPS energies. There hadronic 
transport models without non-standard modifications 
clearly fail to describe the measured $\Xi$ yields. 
We find that meson-baryon reactions not only between anti-Kaons 
and hyperons but also with non-strange particles lead
to a substantial production of $\Xi^-$ in Au+Au collisions around $6~A$ GeV.
Within the present model, the $\Xi/(\Lambda+\Sigma^0)$ ratio is proportional
to the energy density. 

\section*{Acknowledgements}

The authors acknowledge stimulating discussions with 
Dr. Helmut Oeschler at Darmstadt University of Technology.
This work used computational resources provided
by the Centre Calcule at Lyon, France, and the Center for
Scientific Computing (CSC) at Frankfurt, Germany.
This work was supported by GSI, BMBF, DFG.

\end{document}